\documentstyle{article}

\font\tenbf=cmbx10
\font\tenrm=cmr10
\font\tenit=cmti10
\font\elevenbf=cmbx10 scaled\magstep 1
\font\elevenrm=cmr10 scaled\magstep 1
\font\elevenit=cmti10 scaled\magstep 1

\font\ninerm=cmr9

\begin{document}
\begin{center}{{\tenbf   W, Z AND HIGGS SCATTERING AT SSC ENERGIES%
\footnote{\ninerm\baselineskip=11pt Presented by James M. Johnson.}\\  }
\vglue 1.0cm
{\tenrm Suraj N. Gupta and James M. Johnson\\}
\baselineskip=13pt
{\tenit Department of Physics, Wayne State University\\}
\baselineskip=12pt
{\tenit Detroit, MI  48202, USA\\}
\vglue 0.3cm
{\tenrm and\\}
\vglue 0.3cm
{\tenrm Wayne W. Repko\\}
{\tenit Department of Physics and Astronomy, Michigan State University\\}
\baselineskip=12pt
{\tenit East Lansing, MI  48824, USA\\}
\vglue 0.8cm
{\tenrm ABSTRACT}}
\end{center}
\vglue 0.3cm
{\rightskip=3pc
 \leftskip=3pc
 \tenrm\baselineskip=12pt
 \noindent
        We examine the scattering of longitudinal $W$, $Z$ and
Higgs bosons in the Standard Model using the equivalent
Goldstone-boson Lagrangian.  Our calculations include the full one-loop
scattering matrix between the states $W^+_LW^-_L$, $Z_LZ_L$ and $HH$
with no restrictions on the relative sizes of $M_H$ and $\sqrt{s}$. In
addition to deriving the perturbative eigen-amplitudes, we also obtain
quite striking results by unitarizing the amplitudes with the use of the
K-matrix and Pad\'e techniques.
\vglue 0.6cm}
{\elevenbf\noindent 1. Introduction}
\vglue 0.2cm
\baselineskip=14pt
\elevenrm
	There has been much recent interest in the scattering of
longitudinal gauge bosons in the Standard Model.
Theoretically, the symmetry breaking mechanism of the Electroweak
theory is not well known, especially in the case of a strongly coupled
symmetry breaking sector.  Therefore it is natural to examine the most
basic processes, such as the scattering of the longitudinal gauge
bosons.  Experimentally, the scattering of gauge bosons will be
measured at future hadron colliders such as the SSC.  For a
sufficiently large Higgs mass, this process will be a main source for
Higgs bosons.  However, for large $M_H$ and therefore strong coupling,
the Feynman amplitudes violate unitarity and must be unitarized.  The
various unitarization methods must include all possible open channels,
and at energies above $2M_H$ the $HH$ channel must be included.
Therefore the whole $3\times 3$ matrix of amplitudes between the states
$W^+_LW^-_L$, $Z_LZ_L$ and $HH$ must be calculated.  Moreover, since we
are interested in strong coupling, the tree amplitude will be
insufficient, and we will need the amplitudes to at least one-loop.
\vglue 0.6cm
{\elevenbf\noindent 2. Calculation}
\vglue 0.4cm
	We performed this calculation with the aid of the Goldstone
Boson Equivalence Theorem, which allows one to replace the
longitudinal vector gauge bosons with the corresponding scalar
goldstone bosons.  Previously, the $3\times 3$ scattering matrix has
been calculated at the tree-level and the $2\times 2$ submatrix of
gauge boson scattering to one-loop.  The one-loop calculation of the
$HH$ channels is considerably more difficult because of the larger
number of diagrams and their mathematical complexity.  The complexity
of diagrams increases with the number of massive internal propagators,
and the new diagrams contained more of these.  For example, of a total
of six box diagrams in the entire calculation, five were new boxes
needed for the $HH$ channels. After adding the contributions of all of
the diagrams, the amplitudes were then numerically integrated to yield
the s-wave projections, shown in Figure~\ref{amps}.
\begin{figure}[p]
\setlength{\unitlength}{1.in}
\begin{picture}(6.,2.2)
\put(0.,-0.1){\framebox(2.,2.){Figure 1a.} }
\put(3.,-0.1){\framebox(2.,2.){Figure 1b.} }
\end{picture}
\caption{Absolute value of the s-wave Feynman amplitudes of
$W^+_LW^-_L\rightarrow HH$
and $HH\rightarrow HH$ scattering for $M_H=$ $500$, $750$ and $1000$~GeV.
\label{amps}}
\end{figure}
\vglue 0.6cm
{\elevenbf\noindent 3. Unitarization}
\vglue 0.4cm
	Although the Feynman expansion is unitary as a whole, unitarity
is violated order-by-order.  This is especially noticeable at the Higgs
pole, where the Feynman amplitude becomes infinite (see
Figure~\ref{unit}a).  One remedy is to add a finite width for the Higgs
boson.  However, this is an {\elevenit ad hoc} solution which will not
solve the problem of unitarity violation due to large coupling.  In
Figure~\ref{unit}b for a $M_H=1000$~GeV, the absolute value of the
Feynman amplitude is larger than one for all energies above the Higgs
pole.

	The solution is to consistently unitarize the amplitudes.  We
have considered two popular methods: the K-matrix and Pad\`e
unitarizations. If one starts with a Feynman expansion: ${\bf A}_1+{\bf
A}_2+\ldots$, where ${\bf A}_1$ is the matrix of tree amplitudes and
${\bf A}_2$ is the matrix of one-loop corrections, then the K-matrix
unitarization is given by $\Re ({\bf A}_1+{\bf A}_2)\left[{\bf I}+i \Re
({\bf A}_1+{\bf A}_2)\right]^{-1}$ and the Pad\'e by ${\bf A}_1\left[
{\bf A}_1-{\bf A}_2\right]^{-1}{\bf A}_1$.

	Since both techniques are given by matrix expressions, it is
not surprising that the channels become mixed, and contributions of
other channels can influence even the
$W^+_LW^-_L\rightarrow W^+_LW^-_L$
scattering amplitude.  Figure~\ref{unit} shows
\begin{figure}[p]
\setlength{\unitlength}{1.in}
\begin{picture}(6.,2.3)
\put(0.,-0.1){\framebox(2.,2.){Figure 2a.} }
\put(3.,-0.1){\framebox(2.,2.){Figure 2b.} }
\end{picture}
\caption{Comparison of the Feynman amplitude for
$W^+_LW^-_L\rightarrow W^+_LW^-$ with
the K-Matrix and Pad\'e unitarized amplitudes. Results are shown for
$M_H=$ $500$ and $1000$~GeV. \label{unit}}
\end{figure}
the effects of these unitarizations on the Feynman amplitudes for this
process.  It is apparent that both techniques nicely unitarize the
Higgs resonance, without the need to put in a width by hand. In the
case of $M_H=1000$~GeV, they both also reduce the large, unitarity
violating amplitude above resonance.  However, there are differences
between them.  In the case of $M_H=500$~GeV, an additional resonance
appears at $\sqrt{s}=2800$~GeV in the Pad\'e unitarization, but not for
the K-matrix.  For $M_H=1000$~GeV the Pad\'e amplitude also shows the
effect of the $HH$ threshold at $\sqrt{s}=2000$~GeV, more so than for
the K-matrix.  How much physical significance to give to the
interesting features from the Pad\'e unitarization is unclear.

	In Figure~\ref{cross} we present cross-sections for
$pp\rightarrow W^+_LW^-_L
\rightarrow W^+_LW^-_LX$ at the SSC for the two unitarizations.
For $M_H=500$~GeV
\begin{figure}[p]
\setlength{\unitlength}{1.in}
\begin{picture}(6.,2.1)
\put(0.,-0.1){\framebox(2.,2.){Figure 3a.} }
\put(3.,-0.1){\framebox(2.,2.){Figure 3b.} }
\end{picture}
\caption{Cross-sections for $pp\rightarrow W^+_LW^-_L\rightarrow W^+_LW^-_L X$
	 at the SSC for $M_H=$ $500$ and $1000$~GeV. The dashed line is
the K-matrix unitarization and the dotted is the Pad\'e unitarization.
\label{cross}}
\end{figure}
there is little difference at experimentally realizable energies, while
for a larger Higgs mass there is a much larger difference between the
unitarizations.
\vglue 0.5cm
{\elevenbf \noindent 5. Acknowledgements \hfil}
\vglue 0.4cm
This work was supported in part by the U.S. Department of Energy under grant
No. DE-FG02-85ER40209 and National Science Foundation grant 90-06117.
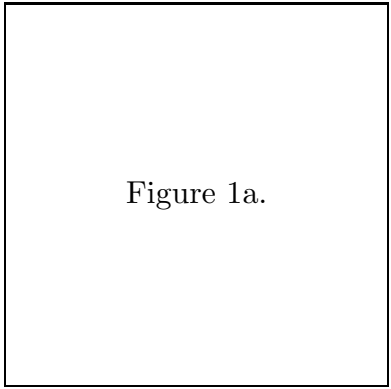
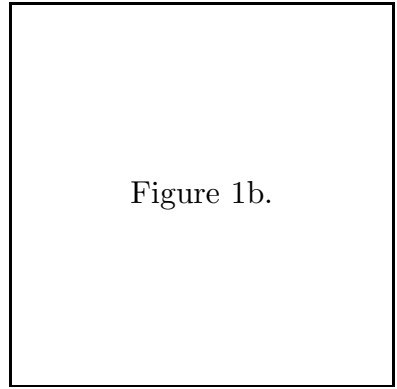
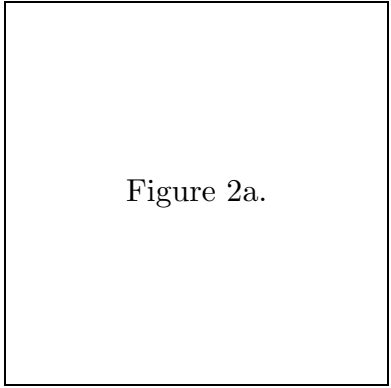
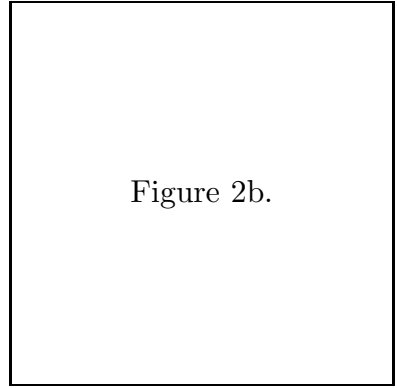
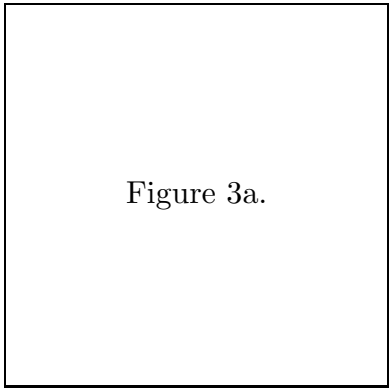
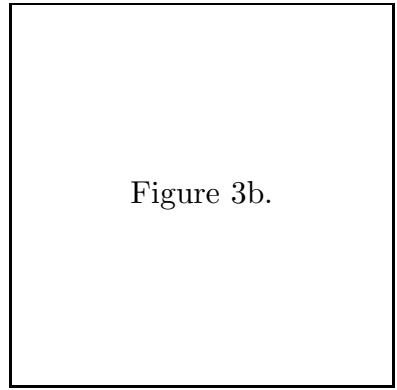
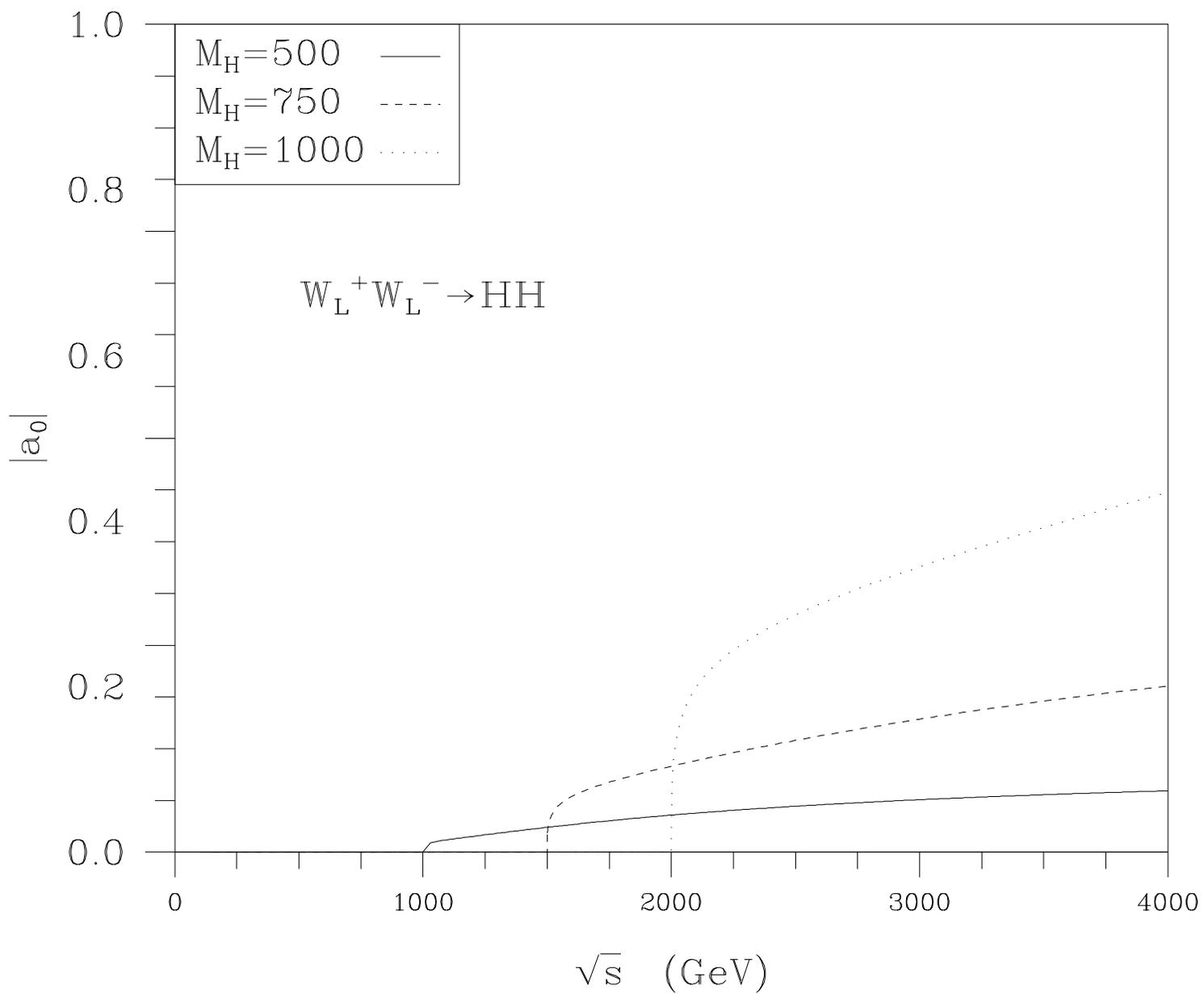
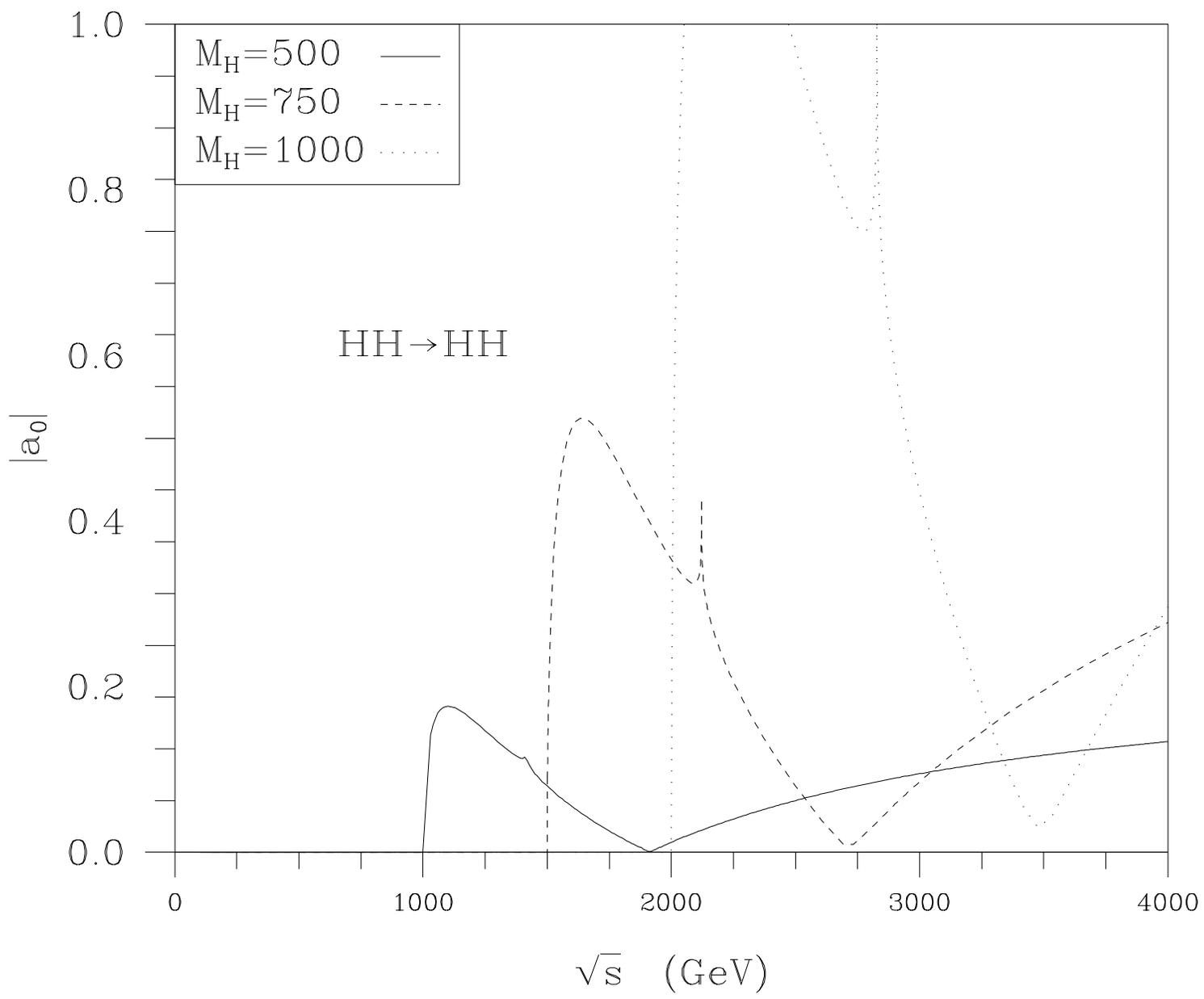
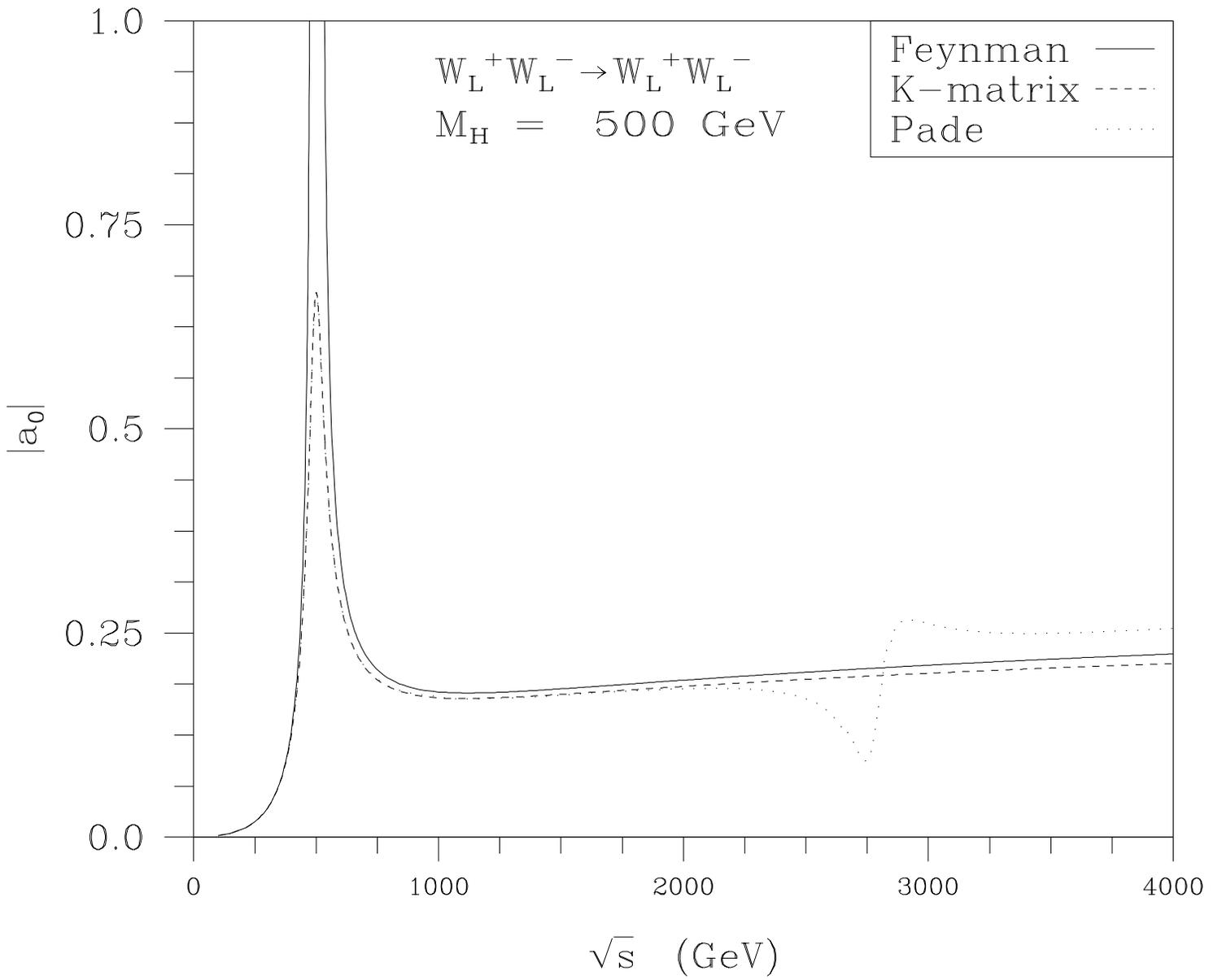
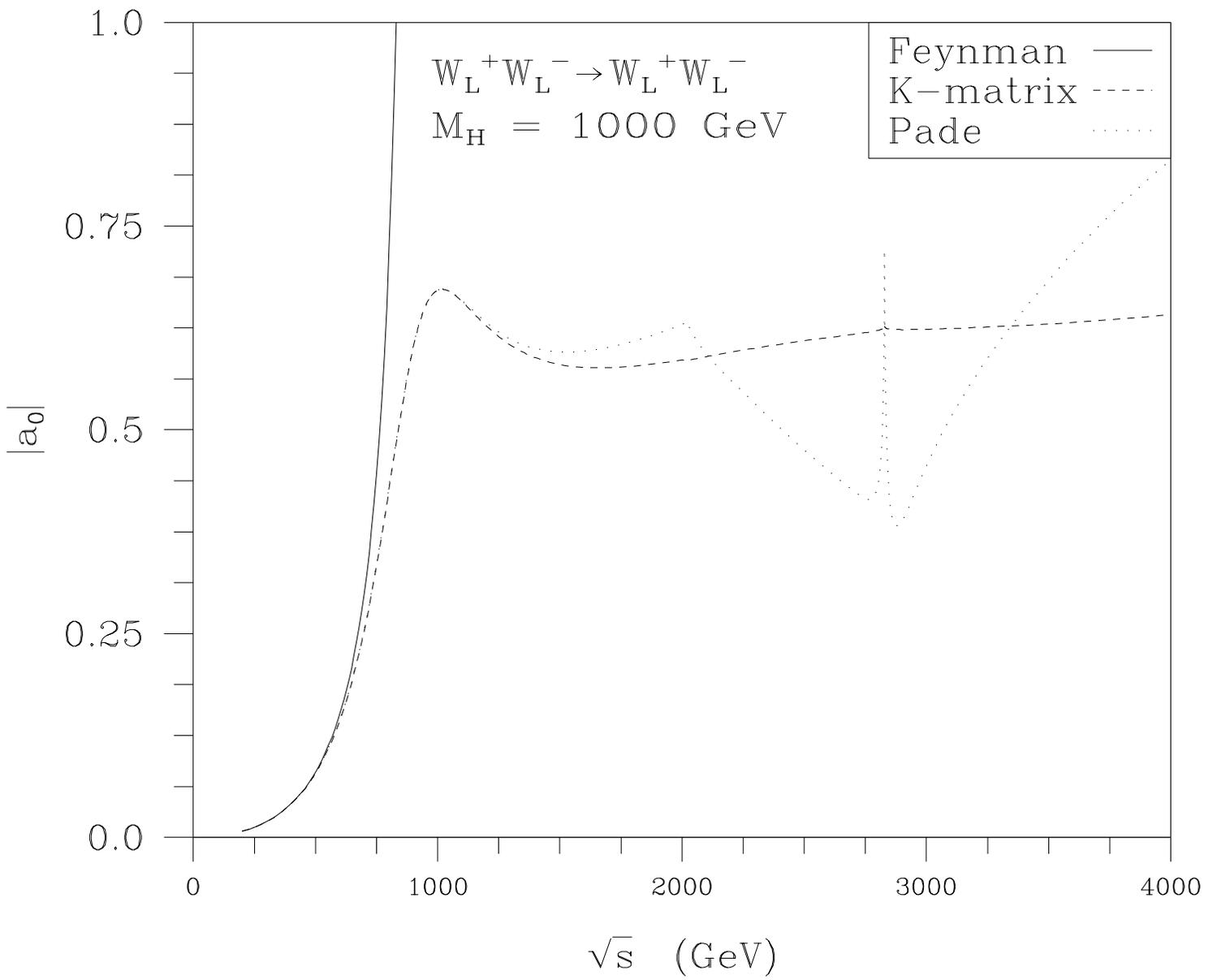
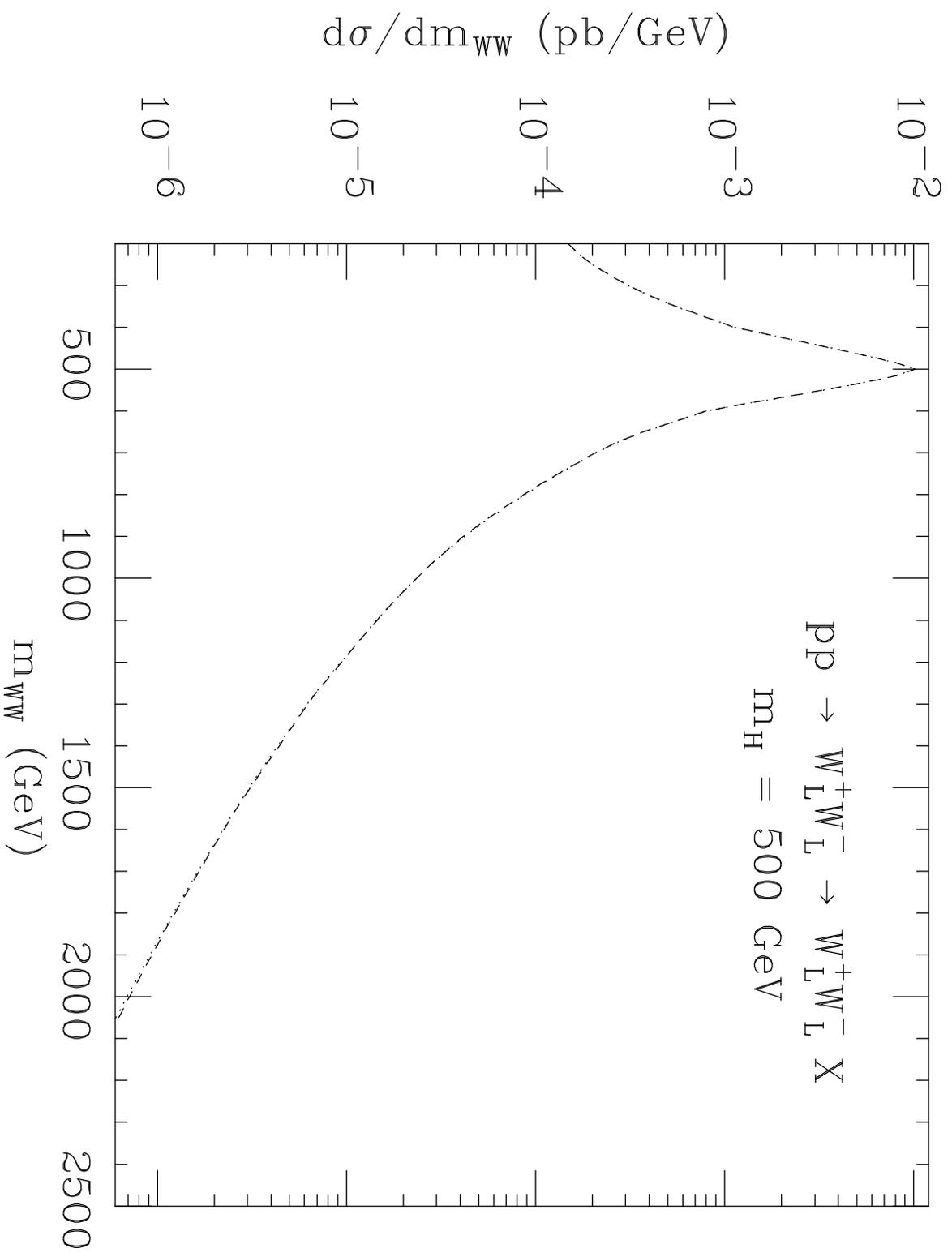
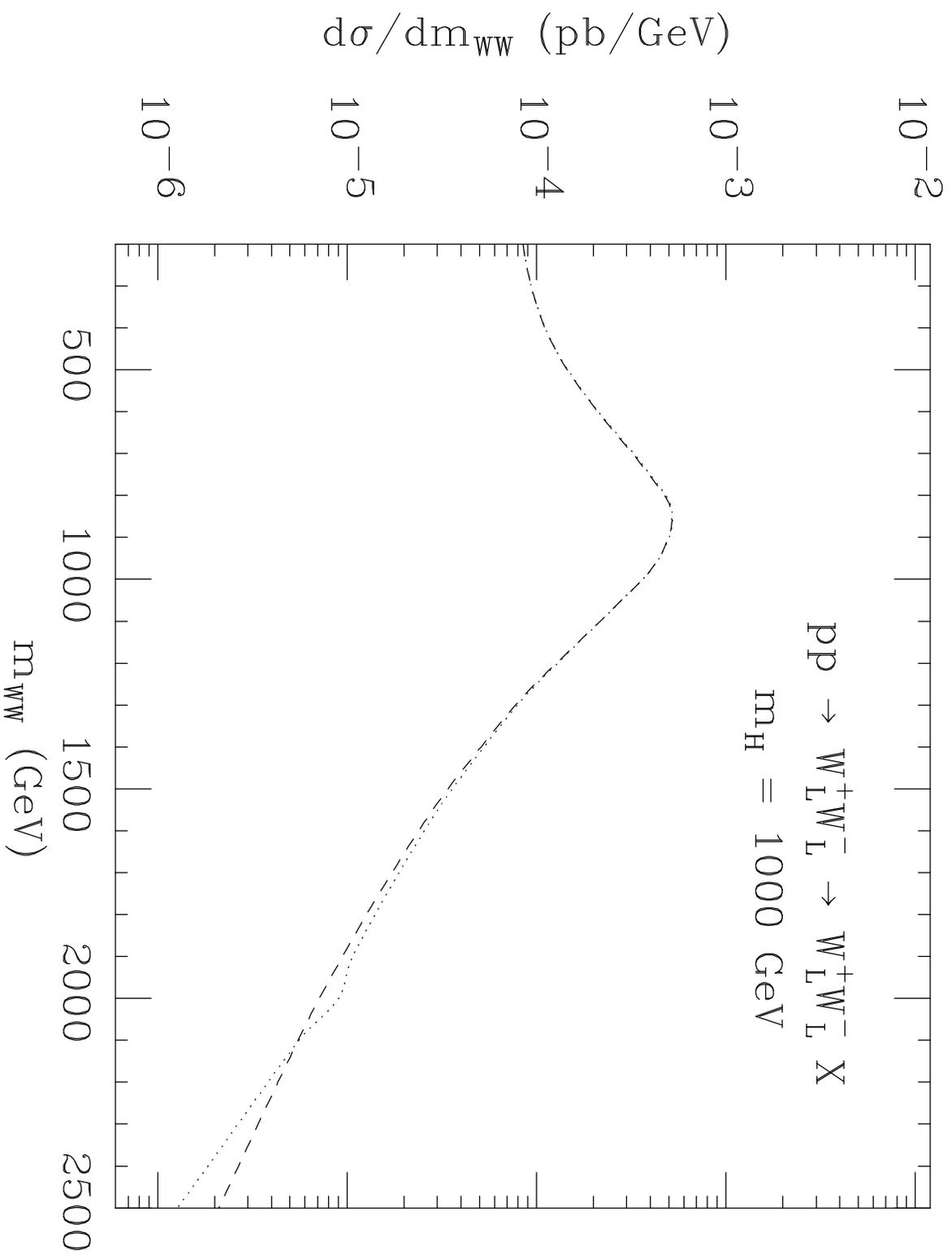
\end{document}